\documentclass[draft]{agujournal2018}
\usepackage{apacite}
\usepackage{url} 
\usepackage{lineno}
\draftfalse

\journalname{JGR: Space Physics}

\begin{document}

\title{3D-modeling of Callisto's Surface Sputtered Exosphere Environment}

\authors{Audrey Vorburger,\affil{1}, Martin Pfleger,\affil{2,3}, Jesper Lindkvist,\affil{4,5}, Mats Holmstr\"{o}m,\affil{4}, Helmut Lammer,\affil{3}, Herbert I. M. Lichtenegger, \affil{3}, Andr\'e Galli,\affil{1}, Martin Rubin,\affil{1}, Peter Wurz\affil{1}}

\affiliation{1}{Physikalisches Institut, University of Bern, Bern, Switzerland.}
\affiliation{2}{Institute for Chemical Engineering and Environmental Technology, Graz, University of Technology, Austria.}
\affiliation{3}{Space Research Institute, Austrian Academy of Sciences, Graz, Austria.}
\affiliation{4}{Swedish Institute of Space Physics, Kiruna, Sweden.}
\affiliation{5}{Department of Physics, Ume\r{a} University, Ume\r{a}, Sweden.}

\correspondingauthor{Audrey Vorburger}{vorburger@space.unibe.ch}

\begin{keypoints}
\item We investigate Callisto's surface sputtered exosphere. Two cases are investigated: One with an ionosphere present, one with no ionosphere.
\item The case where no ionosphere is present produces an asymmetric exosphere, with favoring of the ram direction (direction of plasma arrival).
\item With an ionosphere present, the exosphere is almost uniform and densities are $\sim$2.5 times lower than in the case without an ionosphere.
\end{keypoints}

\begin{abstract}
We study the release of various elements from Callisto's surface into its exosphere by plasma sputtering. The cold Jovian plasma is simulated with a 3D plasma--planetary interaction hybrid model, which produces 2D surface precipitation maps for magnetospheric H$^+$, O$^+$, O$^{++}$, and S$^{++}$. For the hot Jovian plasma, we assume isotropic precipitation onto the complete spherical surface. Two scenarios are investigated: One where no ionospheric shielding takes place and accordingly full plasma penetration is implemented (\textit{`no ionosphere'} scenario), and one where an ionosphere lets virtually none of the cold plasma but all of the hot plasma reach Callisto's surface (\textit{`ionosphere'} scenario). In the 3D exosphere model, neutral particles are sputtered from the surface and followed on their individual trajectories. The 3D density profiles show that whereas in the  \textit{`no ionosphere'} scenario the ram direction is favored, the  \textit{`ionosphere'} scenario produces almost uniform density profiles. In addition, the density profiles in the \textit{`ionosphere'} scenario are reduced by a factor of $\sim$2.5 with respect to the \textit{`no ionosphere'} scenario. We find that the Neutral gas and Ion Mass spectrometer, which is part of the Particle Environment Package on board the JUICE mission, will be able to detect the different sputter populations from Callisto's icy surface and the major sputter populations from Callisto's non-icy surface. The chemical composition of Callisto's exosphere can be directly linked to the chemical composition of its surface, and will offer us information not only on Callisto's formation scenario but also on the building blocks of the Jupiter system.

\end{abstract}

\section{Introduction}
It is well established that Callisto's surface consists of both ice and mineral components (\citet{Greeley2000}, \citet{Kuskov2005}), though what their volume mixing ratio is and to what degree the moon is differentiated still remains unclear (e.g., \citet{Nagel2004} and references therein).

The chemical ice composition of Jupiter's large satellites depends mainly on the early thermodynamic conditions within the Jovian environment prior to their accretion (\citet{Mousis2006}; \citet{Alibert2005a}; \citet{Alibert2005b}). Two formation scenarios that result in two distinct ice surface compositions are often associated with the icy moons. One of these compositions represents an oxidizing state, which is based on the assumption that the building blocks of the Galilean satellites were formed in the protosolar nebula. The second composition represents a reducing state, based on the assumption that the satellites accreted from building blocks that condensed in Jupiter's sub-nebula. Measuring the chemical ice compositions of the Jovian satellites would bring important constraints on both their formation conditions and on the thermodynamic conditions that took place in the protosolar nebula during the formation of the giant planets. ESA's JUpiter ICy moons Explorer (JUICE) \citep{Grasset2013} with its various instruments will help us to distinguish between the different formation scenarios. 

To date, only CO$_2$ was directly measured in Callisto's exosphere \citep{Carlson1999}. Other species, including CO, O$_2$, O, and C, have been inferred indirectly from measurements or have been modeled to match measurements (e.g., \citet{Kliore2002, Strobel2002, Cunningham2015}). We thus still know only very little about Callisto's atmosphere.

Concerning surface composition, two distinct absorption bands in Callisto's reflectance spectra were attributed to the presence of CO$_2$ and SO$_2$ molecules (\citet{Carlson1996}; \citet{McCord1998}). Both species seem to be distributed asymmetrically over the surface of Callisto, where the trailing hemisphere (sub-plasma hemisphere) is more enriched in CO$_2$ and the leading hemisphere (plasma opposing hemisphere) is more enriched in SO$_2$ \citep{Hibbitts2000}). Atmospheric CO$_2$ column densities at the dayside have been estimated by \citet{Carlson1999} to be about $8\times 10^{14}$ cm$^{-2}$. \citet{Calvin1991} presented a compilation of spectral observations of Callisto's surface from 0.2 to 4 $\mu$m, exhibiting properties that are similar to those of phyllosilicate minerals commonly found in carbonaceous chondrite meteorites. For that reason Cl chondrites are often chosen to represent Callisto's mineral component, matching these observations well.

Upper hybrid resonance measurements obtained by the plasma wave instrument on board of Galileo revealed peak electron densities near Callisto's orbit of about 400 cm$^{-3}$ \citep{Gurnett2000}. Electron measurements were also carried out by the Galileo radio occultation instrument \citep{Kliore2002}, but their derived electron densities could not be reproduced solely by ionization of the observed CO$_2$ atmosphere \citep{Carlson1999}. Therefore, \citet{Kliore2002} proposed an O$_2$ component in addition to a tenuous CO$_2$ gas envelope with an inferred O$_2$ column density of (3--4)$\times 10^{16}$ cm$^{-2}$. Recently, \citet{Cunningham2015} inferred from observations of doublet and triplet lines of atomic oxygen the existence of an O$_2$-dominated ``atmosphere'' related to Callisto's leading/Jupiter-facing hemisphere by using the far-UV sensitivity of the Cosmic Origins Spectrograph (COS) on board of the Hubble Space Telescope (HST). From the HST observations these authors infer a leading hemisphere extended molecular oxygen corona with an O$_2$ column density of 3.4$\times 10^{15}$ cm$^{-2}$ \citep{Cunningham2015}. These authors mention that the highly variable O$_2$ density could be an order of magnitude larger at the trailing hemisphere which does agree well with Chapman-type ionospheric layers and inferred column densities obtained by \citet{Kliore2002} for the Callisto C22 and C23 flyby's. Furthermore, the aperture-filling emissions implied that there is also an extended atomic oxygen corona present with an exobase density of up to $\approx 10^4$ cm$^{-3}$ (where exobase denotes the base altitude of the exosphere). These findings are consistent with measurements of the atomic oxygen surface column density for which an upper limit of about 2.5$\times 10^{13}$ cm$^{-2}$ was found \citep{Strobel2002}.

Depending on the plasma environment in Jupiter's magnetosphere, environmental conditions and geographical surface areas on Callisto, the exobase altitude may be located at the satellite's surface or may move atop a collision dominated atmosphere envelope tens kilometers above the surface. That this variation exists can also be inferred from Galileo radio occultation measurements \citep{Kliore2002}. It was found that Callisto can have a substantial ionosphere (i.e., C22 and C23 flyby) with a Chapman-type layer when the trailing hemisphere (which correlates with the magnetospheric ram-side) is illuminated by sunlight \citep{Kliore2002}. The C20 flyby, which had a similar configuration as C22 and C23, though, yielded electron densities three to four times lower. This lower electron density indicates that during flyby C20 the density of the neutral gas envelope was lower than during the flybys C22 and C23. During the ram-side / nightside flyby C9, where the leading hemisphere was illuminated by sunlight, no ionosphere was discernible \citep{Kliore2002}. This led \citet{Kliore2002} to speculate that impact of energetic ions from the magnetosphere of Jupiter and photo-ionization by the sunlight might be needed to create an ionosphere. This disagrees with the C3 and C10 flybys, though, where Callisto's position with respect to Jupiter and the Sun was almost identical to the C9 flyby, but where the flyby was on Callisto's dayside rather than on the nightside. During both flybys a factor 100--1000 increase in plasma density was observed \citep{Gurnett1997, Gurnett2000}. 

The Neutral and Ion Mass spectrometer (NIM), which is part of the Particle Environment Package (PEP) \citep{Barabash2013} on board JUICE, will be able to shed light onto Callisto's formation scenario and provide information on the building blocks of the Jupiter system by measuring the neutral and ion composition of Callisto's (and the other Galilean moons') exosphere (see also \citet{Vorburger2015}). NIM is able to determine the exospheric chemical composition with high mass resolution and unprecedented sensitivity: The mass resolution is $M/\Delta M >$ 1100 in the mass range 1--1000 u and NIM's energy range is 0 -- 5 eV for neutrals and $<$10 eV for ions. The detection level for neutral gas is $1 \times 10^{-16}$ mbar for a 5-second accumulation time, which corresponds to a particle density of about 1 cm$^{-3}$. 

Under the assumption that sputtering releases all species present on the surface stoichiometrically into the exosphere \citep{Plainaki2018}, this process allows us to derive the chemical composition of the surface of Callisto from NIM measurements during the flybys. In the present study we thus focus on species that are released from Callisto's surface by plasma sputtering and investigate with a 3D model how the different species are distributed geographically around the satellite.

In Sect.\ 2 we present the applied 3D hybrid plasma interaction model that produces ion precipitation maps, the implemented ice and mineral surface compositions, the sputter yield computations, and the 3D exosphere model. In Sect.\ 3 we present the results of our simulations for two different scenarios: One where no ionosphere is present and the magnetospheric plasma reaches Callisto's surface almost unhindered (called \textit{`no ionosphere'} scenario), and one where an ionosphere shields the surface almost completely from the thermal magnetospheric plasma and only energetic ions (also refereed to as radiation) reach the surface (called \textit{`ionosphere'} scenario). In Sect.\ 4 we discuss the relevance of our results for the planned PEP/NIM measurements during the JUICE mission. The paper ends with a brief conclusion given in Sect.\ 5.

\section{Description of plasma and exosphere model}
The aim of this work is to model Callisto's surface sputtered exosphere in 3D. First, we apply a 3D plasma-planetary-interaction hybrid model that produces 2D ion precipitation maps of magnetospheric H$^+$, O$^+$, O$^{++}$, and S$^{++}$ onto the satellite's surface. The obtained 2D precipitation maps are then used to calculate the corresponding latitude and longitude dependent sputter fluxes by multiplying the precipitation maps with the energy dependent sputter yields. This result in turn serves as an input for the exosphere model, where we compute for each surface species the resulting exospheric density profile in 3D. The different models and assumptions are described in the subsections below.

\subsection{Coordinate system and simulation domain}
The coordinate system we use is called ``Satellite Centered Phi-B Coordinates'' (SPhiB), a coordinate system often used to model magnetic field interactions with satellites. The coordinate-system is centered in the middle of Callisto. The $+x$ axis is in the direction of the plasma flow. The $z$ axis is defined such that the $x-z$ plane contains the Jovian magnetic field \textbf{\textit{B}}$_0$. The $y$ axis is normal to \textbf{\textit{B}}$_0$ and $x$, i.e., completes the right-hand system. Note that depending on the location of Callisto with respect to Jupiter's magnetic equatorial plane, the magnetic field might point anywhere from perpendicular to the magnetic equatorial plane to towards Jupiter. Since we are interested in plasma interaction with the satellite, though, in this study Callisto is located in the center of the plasma sheet and the magnetic field points parallel to Jupiter's spin axis. The convective electric field is given by \textbf{\textit{E}}$_0$ = -\textbf{\textit{u}} $\times$ \textbf{\textit{B}}$_0$, where \textbf{\textit{u}} is the plasma bulk velocity. Note that in this configuration the undisturbed convective electric field will always be in the $-y$ direction.

The simulation domain is divided into a Cartesian grid with cubic cells of the size $\Delta x = r_{\rm C}/16$, where $r_{\rm C}$ is the radius of Callisto, i.e., 2410.3~km. We use 9 macro-particles per cell as the initial condition, and include periodic boundary conditions. 

When the simulation is initiated, the ions (macro-particles) are evenly distributed everywhere outside Callisto, according to a Maxwellian distribution. Also, the magnetic field is homogeneous everywhere, and equal to the given external field. We model Callisto as an inert object; all particles are absorbed by the surface.

\subsection{Hybrid plasma interaction model}
The hybrid plasma interaction model used herein has previously been applied to solar wind interaction studies with the Moon \citep{Holmstrom2012} and to plasma interaction studies with Callisto \citep{Lindkvist2015}. In this study we use hybrid simulations similar to \citet{Lindkvist2015} where Callisto is modeled as a conductive body without atmosphere and without ionosphere. In the hybrid approximation, ions are treated as particles and electrons are treated as a massless fluid. Figure~1 shows the flow of the magnetospheric plasma in the hybrid model, where the total ion densities are shown in panels a and b and the mean ion velocity is shown in panels c and d for both the XY and XZ planes. The results are taken after 110~s of simulation time. Further details on the hybrid model used here (including the discretization) can be found in \citet{Holmstrom2011a}; \citet{Holmstrom2011b}; \citet{Holmstrom2012}. In this study, the hybrid plasma interaction model has only been used to model the cold, thermal part of the magnetospheric plasma. We produce individual surface precipitation maps for each ion species of the cold plasma, which co-rotates with Jupiter's magnetic field and most of which arrives from Callisto's ram direction.

\begin{figure}
\includegraphics[width=0.95\columnwidth]{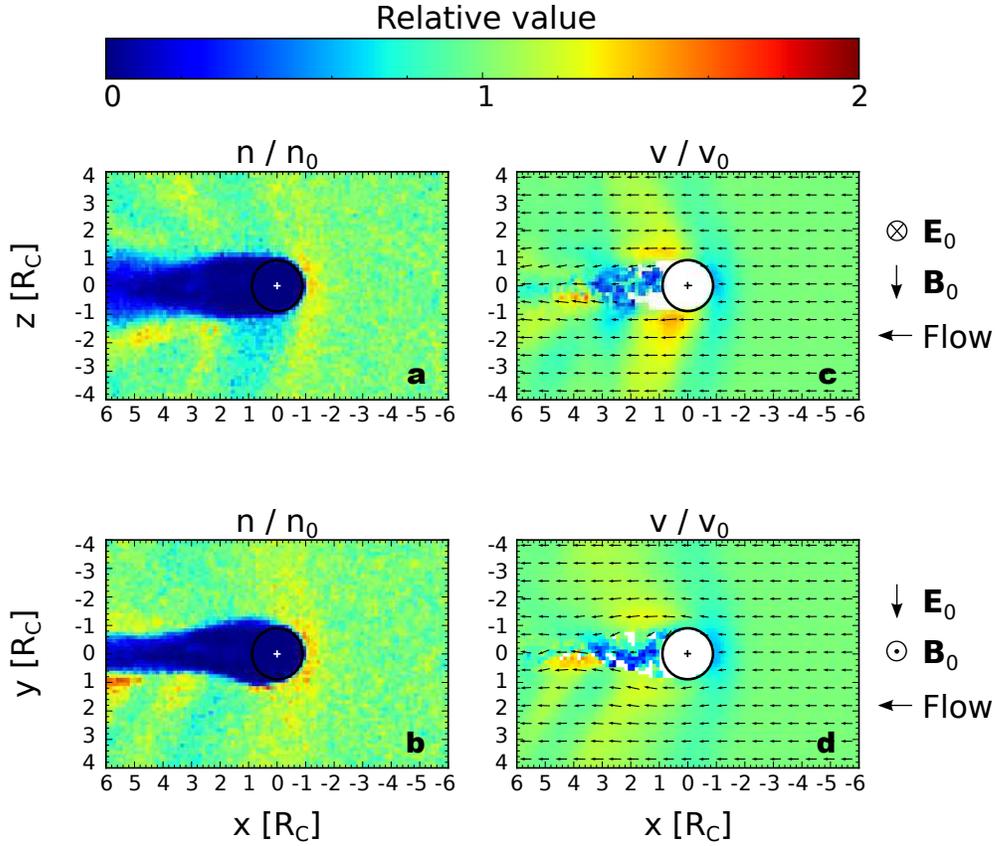}
\caption{Magnetospheric ion flow around Callisto in the hybrid model for two geometric planes: XZ for panels a and c, and XY for panels b and d. Shown in color is the total ion number densities (a, b) and the mean ion speed (c, d). The vector field in (c, d) shows the direction of the mean ion velocity. The axes are given in Callisto radii, $R_\mathrm{C} = 2410.3$~km.}
\end{figure}


In the wake of Callisto to the Jovian magnetospheric plasma flow, the plasma densities of the cold, thermal plasma can be low, or even zero. In such regions of low ion charge density the hybrid solver can have numerical problems due to large gradients in the electric field leading to high accelerations of ions resulting in numerical instabilities. We handle such regions of low ion charge density by solving a magnetic diffusion equation. The magnetic diffusion is also solved inside the obstacle to the plasma flow, in this case inside the spherical inner boundary of the moon's surface \citep{Holmstrom2013}.

For the hot, energetic plasma, which, due to the particles' large gyro-radii, is almost isotropic, we implement uniform plasma arrival onto the complete spherical surface as described below.

\subsection{Plasma parameters}
Jupiter's magnetospheric plasma comprises two populations: The cold, thermal plasma and the hot, energetic plasma (sometimes also referred to as radiation). 

The flow of the cold Jovian magnetospheric plasma passing Callisto at $\approx$~27~R$_\mathrm{J}$ (with R$_\mathrm{J}$ being Jupiter's radius or 69,911~km) is sub-Alfv\'{e}nic or almost sub-Alfv\'{e}nic \citep{Neubauer1998}. The total magnetospheric plasma density at Callisto's orbit fluctuates between $\approx$~0.01 and 1 cm$^{-3}$ (\citet{Neubauer1998}; \citet{Kivelson2004}). We assume for the ambient magnetospheric plasma parameters at Callisto's orbit an average number density of 0.1 cm$^{-3}$ \citep{Kivelson2004}, a bulk ion velocity of $v_{\rm ion}\approx 192$~km/s \citep{Kivelson2004}, an ion temperature of $T_{\rm ion} = 200$~eV \citep{Neubauer1998}, and an electron temperature of $T_{\rm ele} = 100$~eV \citep{Neubauer1998}. The corresponding gyro-radii amount to $\sim$(60 -- 900)~km, depending on the particle mass and charge. We simulated 1D sputtered density profiles within an ion bulk velocity ranging between 132 km/s to 252 km/s and found that the used average bulk velocity resembles the complete velocity range well. The most abundant magnetospheric ion at Callisto's orbit is O$^+$ (\citet{McNutt1981}; \citet{Bagenal1992}). \citet{Bagenal2015} presented plasma conditions at Europa's orbit. We assume that the plasma composition is similar at Callisto's orbit, and implement ion abundance ratios for O$^{+}:$ S$^{++}:$ H$^+:$ O$^{++}$ as $15:10.5:6:4$, with oxygen clearly dominating.

The energetic population of the Jovian plasma is presented in Figure~3 of \citet{Cooper2001}. The hot plasma particles' energy ranges from 10~keV to 100~MeV, with a characteristic energy of $\approx$100~keV  and a temperature of $\sim$20~keV. This results in gyro-radii between 1.3$\times10^3$~km and 5.2$\times10^3$~km, values of the same order as Callisto's radius ($\sim$2410~km). Peak fluxes are on the order of 10$^{10}$~m$^{-2}$~s$^{-1}$. Integrated over energy and solid angle, and considering shielding by the moon, we obtain energetic particle fluxes onto the surface of $\sim2\times 10^6$ cm$^{-2}$ s$^{-1}$, almost the exact same value as the one obtained for the cold population. We implemented the relative abundance of H$^{+}$, O$^{n+}$, and S$^{n+}$ according to Figure~3 in \citet{Cooper2001} as O$^{n+}:$ S$^{n+}:$ H$^+$ = $6:3:1$. Note that these abundances were derived from flux averages over five hours as Callisto moved in and out of the local magnetospheric current sheet and could be variable on shorter time scales, e.g., as Callisto crosses through the center of the current sheet.

The induced magnetic dipole in the simulation has the same magnitude found in the C3 flyby made by the Galileo spacecraft, which is discussed in \citet{Lindkvist2015}, but with an orientation directly opposing the background magnetic field ($+\hat{z}$). The Jovian magnetic field vector is \textbf{\textit{B}}$_0 = -35 \hat{z}$~nT. Because the precipitating ion flux on Callisto's surface is directly connected to the flux of the sputtered species, the ion flux regulates the sputtered density profile. Therefore, one can use the modeled density profiles and adjust them to different ion flux values.

\subsection{Ionospheric Shielding}
Observations show that Callisto has a highly variable ionosphere, as discussed in Section~1. \citet{Strobel2002} implemented a cylindrical symmetric ionosphere and computed that less than 0.07\% of the impinging cold plasma will reach the surface. In a simulation of Io's interaction with the Io plasma torus \citet{Saur1999} showed that as the plasma is diverted it is also slowed down. Assuming that the cold plasma is slowed down by the same degree as it is shielded, the cold plasma bulk velocity decreases from $\approx 192$~km/s to $\approx 134$~m/s when an ionosphere of the given density is present. With water being bound to icy surfaces with a binding energy of $\sim$0.45~eV \citep{Fama2008}, the corresponding ion kinetic energy of 0.0001~eV to 0.003~eV is not high enough to liberate icy surface water molecules anymore. When we thus study sputtering in the case of an ionosphere being present, we only implement sputtering by hot plasma ions but not by cold plasma ions. Note, though, that whereas the slowed cold plasma particles do not contribute to the sputter yield anymore, they are (to 0.07\% of the upstream value) still available for atmospheric interaction processes (i.e., loss processes; see below).

We decided to present two cases in accordance to the Galileo observations during the C3 and C10 versus the C9 flybys discussed above: One labeled \textit{`ionosphere'}, where an ionosphere lets all of the hot plasma but only 0.07\% of the cold plasma pass, and where no cold plasma sputtering yet cold plasma interaction takes place, and one labeled \textit{`no ionosphere'}, where no ionospheric shielding takes place.

\subsection{Magnetospheric ion precipitation maps}
In the hybrid model, the ion number density and mean ion fluxes are stored to a grid with the given cell-size. We define the precipitating ion flux as the component of the mean ion flux that is anti-parallel to the surface normal. Then we choose points on the surface separated by an angle in both longitude and latitude, given by the angular size of half a cell at the equator. At each point on the surface, we interpolate a value for the chosen parameter (ion number density or precipitating flux) using the cells around it. This is done by creating an imaginary cell with its center at the point of interest, and then sum the contributions from the enclosed simulation cells. Since the cell-size, $\Delta x = r_C / 16$, is not infinitely small, the interpolation will result in a value at a mean altitude of $\approx$~36~km. 

Figure~2 shows the impinging fluxes of cold H$^{+}$, O$^{+}$, O$^{++}$, and S$^{++}$ used in the \textit{`no ionosphere'} case. One can see that the flux is higher for O ions than for S and H ions. The majority of magnetospheric ions precipitate onto the surface over the ram side, which is the trailing hemisphere that faces Jupiter's co-rotating magnetospheric plasma flow. In addition, the anti-ram side also experiences some ion fluxes, but the flux values are much lower than on the ram side. Fluxes of ions precipitating onto the surface as function of latitude and longitude are used as input parameters for the surface interaction and sputter yield modeling described below.

\begin{figure}
\includegraphics[width=0.95\columnwidth]{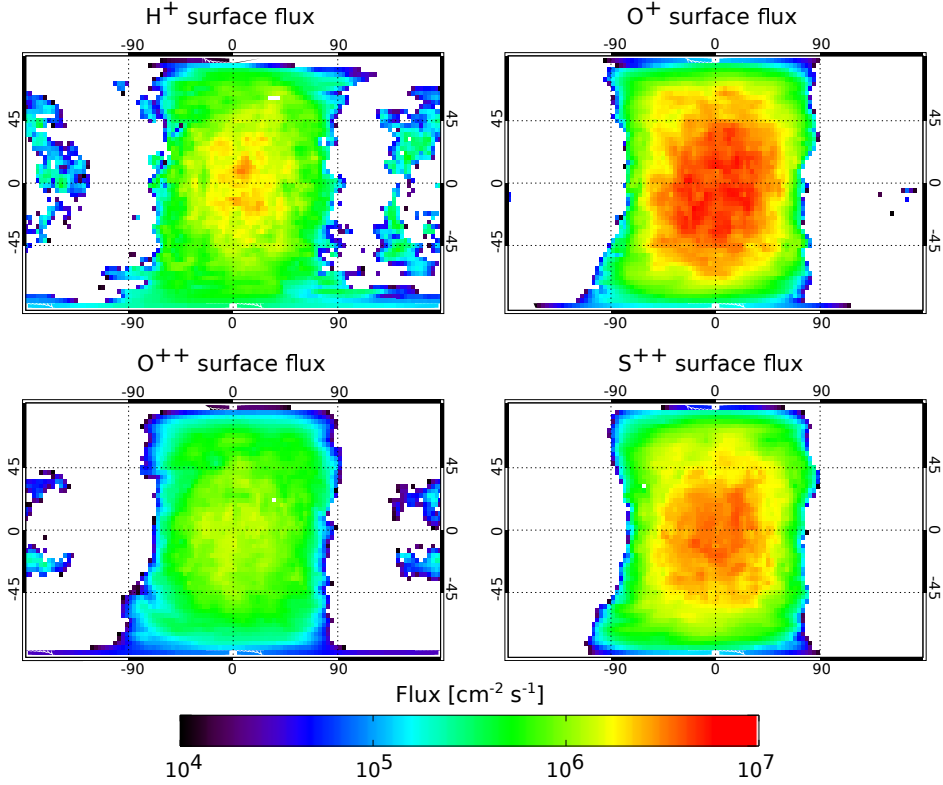}
\caption{Maps of precipitating cold magnetospheric H$^{+}$, O$^{+}$, O$^{++}$, and S$^{++}$ ions onto Callisto's surface in units of cm$^{-2}$ s$^{-1}$ used in the \textit{`no ionosphere'} case. The hemisphere facing Jupiter's co-rotating magnetospheric plasma flow (ram side) lies in the center of the precipitation maps, positive latitudes correspond to Callisto's northern hemisphere, positive longitudes represent the eastern hemisphere.}
\end{figure}

\subsection{Surface composition and sputter modeling}
For this study we assume similar surface compositions and use the ion velocity dependent sputter yields as presented in \citet{Vorburger2015}. 

In our model runs we assume for the icy surface composition the aforementioned oxidizing state, which is based on the assumption that the building blocks of Callisto were formed in the protosolar nebula. The composition belonging to the oxidizing state can be found in Table 1 of \citet{Vorburger2015}. The advantage of taking the oxidizing state rather than the reducing state is that it includes all major volatile compounds. A comparison with the reducing case is simple, since one can just ignore the CO$_x$H$_y$ molecules and instead increase the CH$_4$ and NH$_3$ fractions. For the non-ice compositions we study the compositions of typical LL chondrites for Mg and of typical CI chondrites for Fe (see Table 2 in \citet{Vorburger2015}).

The sputter yields for ices have been calculated based on the experimental results from \citet{Shi1995} and the extension by \citet{Fama2008} and \citet{Galli2018}. For both sputtering from icy and non-icy surfaces we assume that the ion charge state does not influence the sputter yield \citep{Galli2017, Muntean2016}. \citet{Shi1995} found that for solar wind protons the sputter yield of water ice is about a factor 10 higher than for sputtering of mineral grains. Thus, similar as in \citet{Vorburger2015}, electronic excitations related to icy surfaces, which results in much higher sputter yields, are considered for icy surfaces. For the sputter yield of O$_2$, which is a radiolysis product of H$_2$O, we used the sputter yield of H$_2$O and scaled it to the abundance of O$_2$ expected to be present in plasma processed water ice. (For more information see Section~3.3.1 and Table~2 in \citet{Vorburger2018}.) Thermal desorption and thermal accommodation has not been considered in this paper, but will be presented in a follow-up paper. The sputter yields for minerals have been determined with the Stopping and Range of Ions in Matter (SRIM) code, which calculates ion interactions with matter (http://www.srim.org; \citet{Ziegler2004}). The SRIM tool uses a Monte-Carlo simulation method for the calculation of the binary collisions between an impacting ion and atoms on the surface for a given chemical composition of the surface. Inputs are the sputter agent's mass, energy, and angle of incidence. For the surface the input parameters are the present atoms' weight percentages, the thickness of the simulated layer and the layer's density.

The total sputter yield $Y_{\rm i}$ of species $i$ is calculated by $Y_{\rm i}=x^{\rm H^+} Y_{\rm i}^{\rm H^+} + x^{\rm O^{n+}} Y_{\rm i}^{\rm O^{n+}} + x^{\rm S^{n+}} Y_{\rm i}^{\rm S^{n+}}$, where $x^{H^+}$, $x^{O^{n+}}$, and $x^{S^{n+}}$ are the fractions of H$^+$, O$^{n+}$, and S$^{n+}$ ions in Jupiter's co-rotating magnetospheric plasma flow, while $Y_{\rm i}^{\rm H^+}$, $Y_{\rm i}^{\rm O^{n+}}$, and $Y_{\rm i}^{\rm S^{n+}}$ are the sputter yields of species $i$ caused by the precipitating ions.

\subsection{3D exosphere modeling}
The sputter yields of the released chemical species are then used as the basic input into the 3D exosphere model. In this model, the trajectories of a large amount of particles are computed independently (i.e., the model is collision-free) ab initio, i.e., starting at the surface. Each sputtered particle starts its trajectory at the exobase (which in this case is assumed to be Callisto's surface), with an energy sampled randomly from the energy distribution function for sputtered particles:

\begin{equation}
 f(E) = \frac{6E_b}{3-8\sqrt{E_b/E_c}}\frac{E}{(E+E_b)^3}\left(1-\sqrt{\frac{E+E_b}{E_c}}\right),
\end{equation}

\noindent where $E_b$ is the surface binding energy, and $E_c$ is the maximum energy that can possibly be transferred in a binary collision between the impacting particle and the surface atom. Typical energies of sputtered particles are on the order of 0.1~eV. The particles' ejection angles are obtained by randomly sampling a cosine function for the polar angle and a uniform distribution for the azimuth angle. 3D trajectories are computed until the particles either leave the calculation domain, get ionized, are fragmented, or fall back to the surface. Particles that return to the surface are assumed to be completely sticking. More details of the 3D exosphere model are given in \citet{Pfleger2015}.

The sputter rate $Q_{\rm sp}$ in units of s$^{-1}$ of the particles that are generated in a surface element is calculated by 

\begin{equation}
Q_{\rm sp}=\int Y \cdot J dA,
\end{equation}
with $A$ the area related to a surface element, $Y$ the sputter yield, and $J$ the impinging plasma flux.

A total of about $10^{6}$ pseudo-particles equally distributed over Callisto's surface are generated, where the surface production rate $Q_{\rm sp}^{\rm p}$ assigned to each pseudo-particle can be written as

\begin{equation}
Q_{\rm sp}^{\rm p}=\frac{Q_{\rm sp}}{N},
\end{equation}
where $N$ is the number of pseudo-particles of the considered surface element. The simulation domain reaches from the surface up to 2.5$\times 10^{4}$ km or 10.37 Callisto radii (half of Callisto's Hill radius). Particles that move beyond this altitude or impinge at the surface are considered as lost.

In addition to atmospheric escape and surface sticking we consider ionization and dissociation as loss processes. Several different particle populations can ionize or dissociate neutral particles in Callisto's exosphere. Ionization is only effective in a limited energy range, usually a few ten eV to a few ten keV. We accordingly only consider ionization by solar photons, cold plasma electrons, cold plasma ions, and ionospheric O$_2^+$ (the most abundant ion in the ionosphere). For dissociation we consider solar photons and cold plasma electrons, the most efficient dissociation agents. 

For loss processes involving photons we use values given at 1~AU by \citet{Huebner1992} and \emph{http://phidrates.space.swri.edu} and scale them to Callisto's distance to the Sun. To determine the atmospheric loss rate by electron and ion impact we multiply the respective particle fluxes with the interaction specific cross sections. The cold plasma electron flux is taken as the product of the electron density and the electron thermal velocity based on $T_{\rm ele} = 100$~eV, whereas the cold plasma ion flux is obtained directly from our hybrid model. Accordingly, the electron flux is the same in the \textit{`no ionosphere'} and in the \textit{`ionosphere'} case, whereas cold plasma ion ionization becomes ineffective in the \textit{`ionosphere'} case due to the ions' low energy. With the neutral O$_2$ flux being the source for the ionospheric O$_2^+$ flux we try to estimate the ionospheric O$_2^+$ flux based on the neutral sputtered O$_2$ flux. To this end we model the neutral O$_2$ flux and multiply it with the ionization fraction, i.e., the number of particles that were ionized during our Monte-Carlo simulations ($\approx$0.04\%). The obtained O$_2^+$ particle flux of $1\times10^{10}$~cm$^{-2}$~s$^{-1}$ is about two orders of magnitude higher than the O$_2^+$ flux that was similarly computed for Europa \citep{Vorburger2018}. This agrees well with the neutral O$_2$ density being also approximately 100 times denser on Callisto than on Europa. The cross sections were assembled from \citet{Garrett1985, Freund1990, Tawara1990, McCallion1992, Kanik1993, Straub1996, Boivin1998, McGrath1998, Deutsch2000, Johnson2002b, Luna2005, Riahi2006}, and \citet{McConkey2008} for an electron energy of $\sim$100~eV and an ion energy of a few keV.


\begin{table}
\caption{Molecular reaction rates for O$_2$ and Mg. `Chex' refers to charge exchange.}
\begin{tabular}{l | c  c  c | c  c  c }
 & \multicolumn{3}{c | }{`no ionosphere'} & \multicolumn{3}{c}{`ionosphere'} \\
process & $\sigma$~(10$^{-16}$cm$^2$) & J~(cm$^{-2}$s$^{-1}$) & r~(10$^{-6}$ s$^{-1}$) & $\sigma$~(10$^{-16}$cm$^2$) & J~(cm$^{-2}$s$^{-1}$) & r~(10$^{-6}$ s$^{-1}$) \\
\hline
  \textbf{O$_2$} & & & & & & \\
  h$\nu$ dissociation	&		&				& 0.15		&		&				& 0.15 		\\
  e$^-$ dissociation	&	0.33	& J$_{e^-}$ = 8.9e07		& 0.0003	&	0.33	& J$_{e^-}$ = 8.9e07		& 0.0003 	\\
  h$\nu$ ionization	&		&				& 0.02		&		&				& 0.02		\\
  e$^-$ ionization	&	1.56	& J$_{e^-}$ = 8.9e07		& 0.014		&	1.56	& J$_{e^-}$ = 8.9e07		& 0.014		\\
  Plasma chex 		&	15	& J$_{i^{n+}}$ = 1.9e06		& 0.003		&		& 				& 		\\
  Ionosph. chex 	&		& 				& 		&	14	& J$_{O_2^+}$ = 1.0e10		& 14.00		\\
  \hline
  Total 	 	&		& 				& 0.19		&		& 				& 14.18		\\
  \hline
  \textbf{Mg} & & & & & & \\
  h$\nu$ ionization	&		&				& 0.02		&		&				& 0.02		\\
  e$^-$ ionization	&	2.4	& J$_{e^-}$ = 8.9e07		& 0.021		&	2.4	& J$_{e^-}$ = 8.9e07		& 0.021		\\
  Plasma chex 		&	$<$20	& J$_{i^{n+}}$ = 1.9e06		& 0.004		&		& 				& 		\\  
  Ionosph. chex 	&		& 				& 		&	86	& J$_{O_2^+}$ = 1.0e10		& 86.00		\\
  \hline
  Total 	 	&		& 				& 0.045		&		& 				& 86.04		\\
\end{tabular}
\end{table}

Table~1 lists ionization and fragmentation loss rates computed for O$_2$ and Mg for the \textit{`no ionosphere'} case and for the \textit{`ionosphere'} case separately. As one can see from this table, if an ionosphere of the calculated density is present, the charge exchange between the ionospheric ions and the neutral particles dominates the neutrals' loss rates by orders of magnitude. If no ionosphere is present, the loss rate is mainly determined by the photon-dissociation rate. We only performed these detailed considerations for neutral O$_2$ and Mg, but use the reported rates for all icy and non-icy species. Cross sections of processes involving other neutral species will be different from the ones reported for O$_2$ and Mg, but on the same order of magnitude. Considering the overall uncertainties in cross sections and particle fluxes, we feel confident that the loss rates reported for O$_2$ and Mg in Table~1 are also applicable for other neutral species.

\section{Results \& Discussion}
Figure~3 shows he density profiles of H$_2$O (the most abundant ice species) sputtered from Callisto's surface for four sensitivity cases (non-realistic scenarios) and two realistic scenarios. The left side panels present the individual contributions of the cold and the hot plasma for a fixed loss rate of $0.2\times10^{-6}$~s$^{-1}$ (\textit{`no ionosphere'} loss rate), the central two panels show the effect of two different loss rates ($0.2\times10^{-6}$~s$^{-1}$ from the \textit{`no ionosphere'} scenario and $14\times10^{-6}$~s$^{-1}$ from the \textit{`ionosphere'} scenario) for full plasma penetration, and right two panels show the resulting total density distributions for the two scenarios \textit{`no ionosphere'} / \textit{`ionosphere'} (with the right plasma penetration and loss rates implemented). Note that the artifacts in the pole directions are a result of the grid discretization.

The comparison between the two plasma populations (left side panels) shows that the cold plasma and the hot plasma contribute about equally to the overall H$_2$O density profiles. The cold plasma creates, as expected, a density profile with a longitude and latitude dependency, where most particles are sputtered in the the direction of plasma arrival, i.e., the trailing side of Callisto. The hot plasma contribution, on the other hand, creates an almost uniform exosphere.

The two different loss rates computed for the two different scenarios do not seem to have a large impact on the density profiles of the neutral exosphere. One does observe a slightly more rapid density decrease in the case of the higher loss rate (e.g., the orange color reaches less far out), but with the ionization rate being small compared to the average particle flight time, the difference in the two profiles is minor.

The density profiles for the two scenarios \textit{`no ionosphere'} / \textit{`ionosphere'} do not show any surprises either. For the \textit{`no ionosphere'} scenario, we implement full plasma penetration and apply the lower loss-rate of $0.2\times10^{-6}$~s$^{-1}$. For the \textit{`ionosphere'} scenario we use 100\% of the hot plasma ions but none of the cold plasma ions for sputtering and apply the higher loss-rate of $14\times10^{-6}$~s$^{-1}$ (i.e., 0.07\% of the cold plasma is still available for charge exchange). The \textit{`no ionosphere'} exosphere density profile contains $\sim$2.5 times as many particles than the \textit{`ionosphere'} profile. The exosphere is less uniform in the \textit{`no ionosphere'} case, with more particles sputtered in the direction of plasma arrival than in other directions (north / south, leading side of Callisto). The \textit{`ionosphere'} density profile is almost identical to the hot plasma density profile, with the only difference being a slightly steeper decrease in density with altitude due to the higher loss rate caused by ionospheric charge exchange processes.

In the following two subsections, we will present the density profiles corresponding to the two scenarios \textit{`no ionosphere'} and \textit{`ionosphere'} separately. 

\begin{figure}[t]
\begin{center}
\includegraphics[width=0.9\columnwidth]{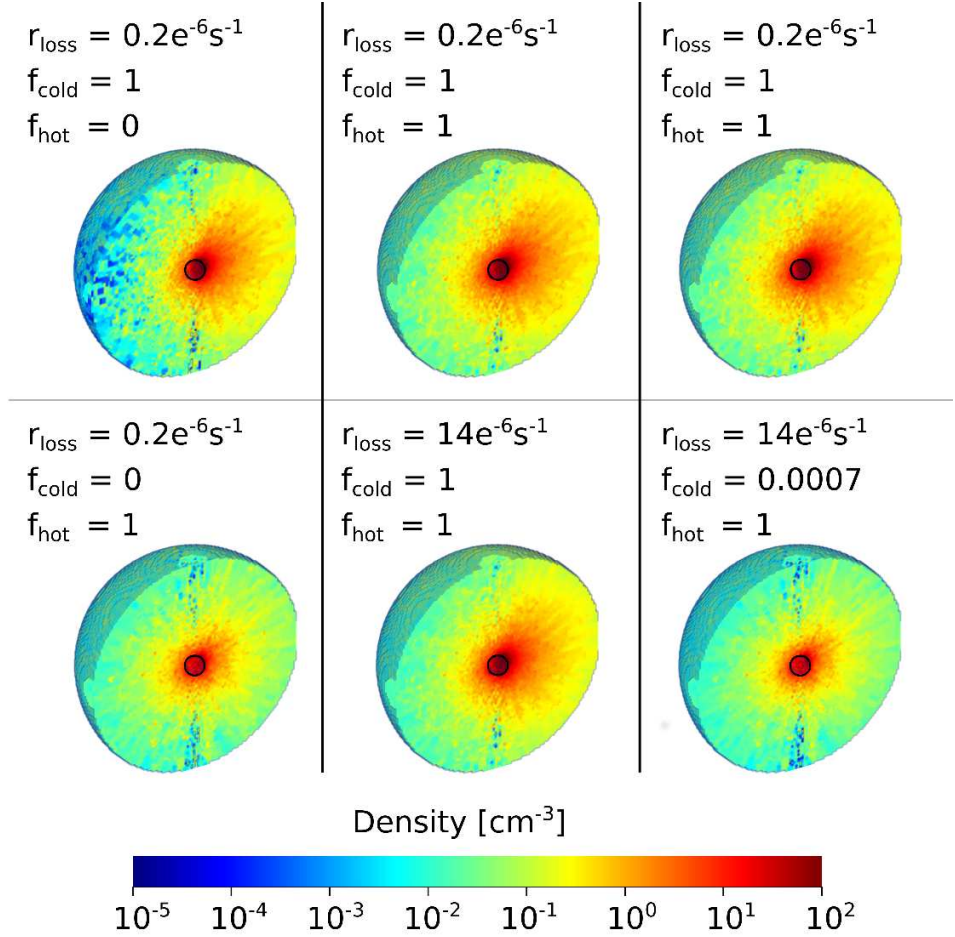}
\caption{Number density cuts in units of cm$^{-3}$ for surface sputtered H$_2$O. The magnetospheric ram side is on the right hand side, and -\textbf{\textit{B}}$_0$ points upwards. The left two panels show the individual sputter contributions by the cold and the hot plasma for a fixed loss rate ($0.2\times10^{-6}$~s$^{-1}$). The middle two panels show the density profiles for full plasma penetration for two different loss rates. The right two panels show the computed H$_2$O density profiles for the \textit{`no ionosphere'} and the \textit{`ionosphere'} cases. The simulation domain extends between the surface and an altitude of 25000 km or 10.37 Callisto radii.}
\end{center}
\end{figure}

\subsection{Without ionosphere}
Figure~4 shows the modeled number density distributions for H$_2$O, CO$_2$, CO, and O$_2$ sputtered from an icy surface (oxidizing case), and for Mg and Fe sputtered from a non-ice surface, with a composition of LL \& CI chondrite, for the scenario where 100\% of both the hot and the cold plasma reach Callisto's surface (\textit{`no ionosphere'}).

The average global number densities of sputtered H$_2$O molecules are $\approx$~110 cm$^{-3}$ near the surface, $\approx$~32 cm$^{-3}$ at 1,000~km, and $\approx$~1 cm$^{-3}$ at 10,000~km. The densities on the ram side hemisphere are $\approx$~1.5 times higher than the global average, whereas the densities on the anti-ram side hemisphere are $\approx$~2 times lower than the global average. The average global number densities of sputtered CO$_2$, CO, and O$_2$ are $\approx$~8, 2, and 5 times lower, respectively, compared to the global H$_2$O densities. The density profiles for the non-ice components are $\approx$~9 times lower for Mg and $\approx$~57 times lower for Fe than for H$_2$O. Those lower non-ice density profiles result from the lower sputter yields for non-ice materials than for icy materials. In addition, Mg is about three times more abundant in LL chondrites than Fe is in CI chondrites and weighs less than half as much as Fe, so the difference in scale height between the two non-ice density profiles is also well understood.

Overall, the density profiles of all sputtered species in the \textit{`no ionosphere'} case exhibit the same characteristics as the density profile of H$_2$O: A denser exosphere on the ram side and less density towards the poles and on the anti-ram side.

The 1D simulations presented in Figure~3 of \citet{Vorburger2015} are $\approx$~3 to 4 times lower than the densities presented in this work. It is noteworthy, that while the model applied in the \citet{Vorburger2015} study is quite similar to the model used herein, there are, besides the dimensionality, two major differences. First of all, \citet{Vorburger2015} only considered the cold plasma distribution, whereas in this work both the cold and the hot plasma contributions are considered. This already explains a factor $\approx$~1.5 difference between the two analyses. Secondly, \citet{Vorburger2015} approximated the cold plasma by three mono-energetic ion fluxes (one for H$^{+}$, one for O$^{+}$, and one for S$^{++}$) while we use the precipitation maps created by our hybrid model. These maps are not only more accurate, but also slightly higher at the sub-plasma point than the averages used in the \citet{Vorburger2015} study. Overall, the herein presented 3D profiles thus agree very well with the 1D profiles computed previously.

\begin{figure}[t]
\begin{center}
\includegraphics[width=0.9\columnwidth]{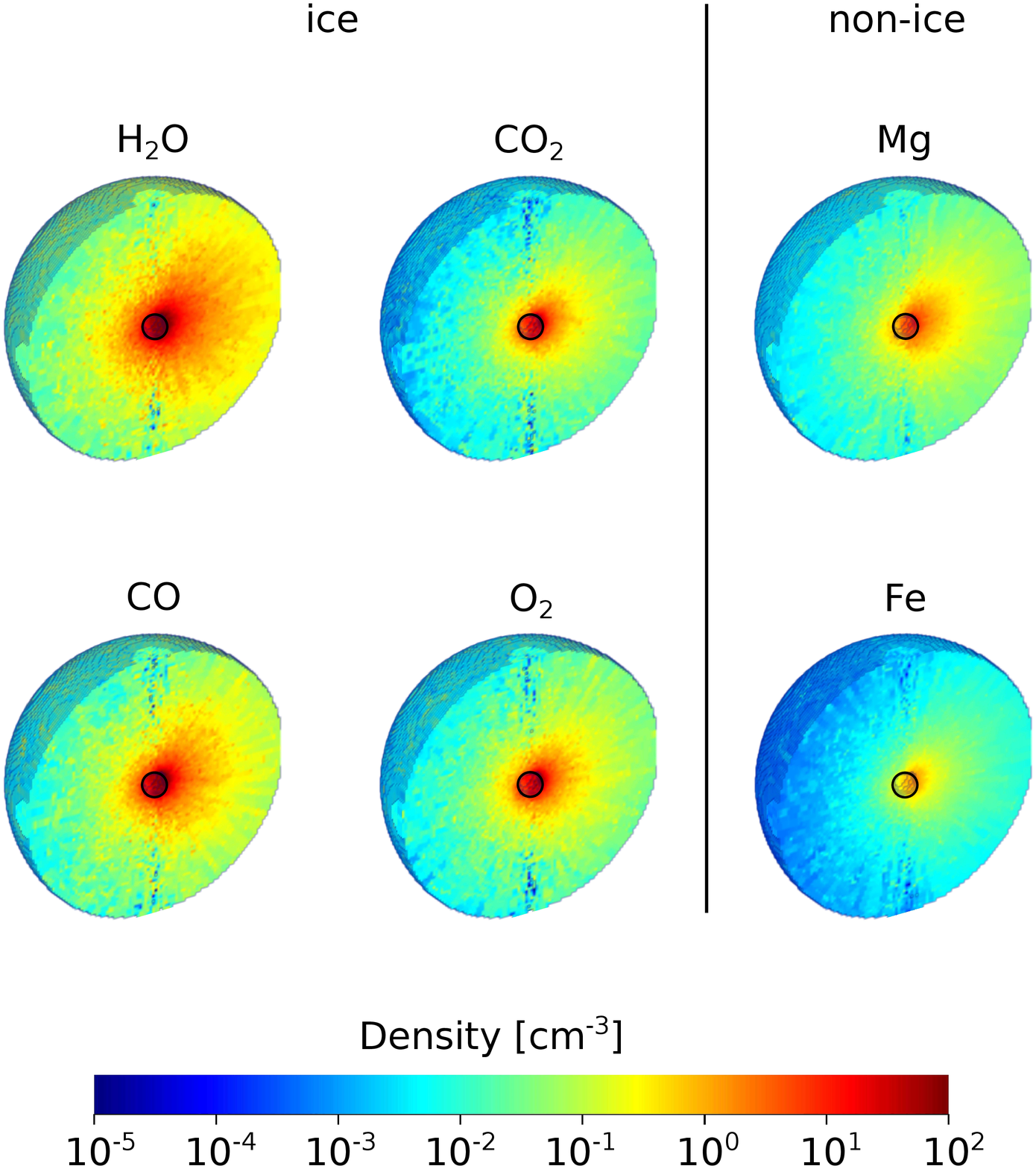}
\caption{Number density cuts in units of cm$^{-3}$ for H$_2$O, CO$_2$, CO, and O$_2$ sputtered from an oxidizing icy surface, Mg sputtered from a LL-type chondritic non-icy surface, and Fe sputtered from a CI chondritic non-icy surface for the \textit{`no ionosphere'} case. The magnetospheric ram side is on the right hand side, and -\textbf{\textit{B}}$_0$ points upwards. The simulation domain extends between the surface and an altitude of 25000 km or 10.37 Callisto radii.}
\end{center}
\end{figure}

\subsection{With ionosphere}
Figure~5 shows the modeled number densities for H$_2$O, CO$_2$, CO, and O$_2$ sputtered from an icy surface (oxidizing case), and for Mg and Fe sputtered from a non-ice surface composition (LL \& CI chondrite) for the scenario where all of the hot plasma but only 0.07\% of the cold plasma ions reaches Callisto's surface (\textit{`ionosphere'}). Remember, that whereas the cold plasma does not contribute to the sputter yield (due to the plasma's low velocity), it is still available for atmospheric loss processes.

The global H$_2$O number densities are $\approx$~76 cm$^{-3}$ near the surface, $\approx$~21 cm$^{-3}$ at 1,000~km, and $\approx$~0.53 cm$^{-3}$ at 10,000~km. They are thus, for the ionospheric density assumed, $\sim$2 times lower than in the \textit{`no ionosphere'} scenario. Note that, as mentioned in the introduction, the ionosphere is highly variable, which makes this factor also highly variable. The average global number densities of sputtered CO$_2$, CO, and O$_2$ are also $\approx$~8, 2, and 5 times lower, respectively. The ratios are thus very similar to the \textit{`no ionosphere'} case. The density profiles for the non-ice components are $\approx$~22 and 108 times lower for Mg and Fe than for H$_2$O. These ratios are larger than the ratios of the \textit{`no ionosphere'} scenario, because mineral sputtering is mainly triggered by cold plasma ions.

Overall, the density profiles associated with the \textit{`ionosphere'} scenario are $\approx$~2.5 times lower for icy constituents and $\approx$5.5 times lower for non-icy constituents than in the \textit{`no ionosphere'} scenario. They are almost uniform, with only a slight favoring of the ram direction. 

\begin{figure}[t]
\begin{center}
\includegraphics[width=0.9\columnwidth]{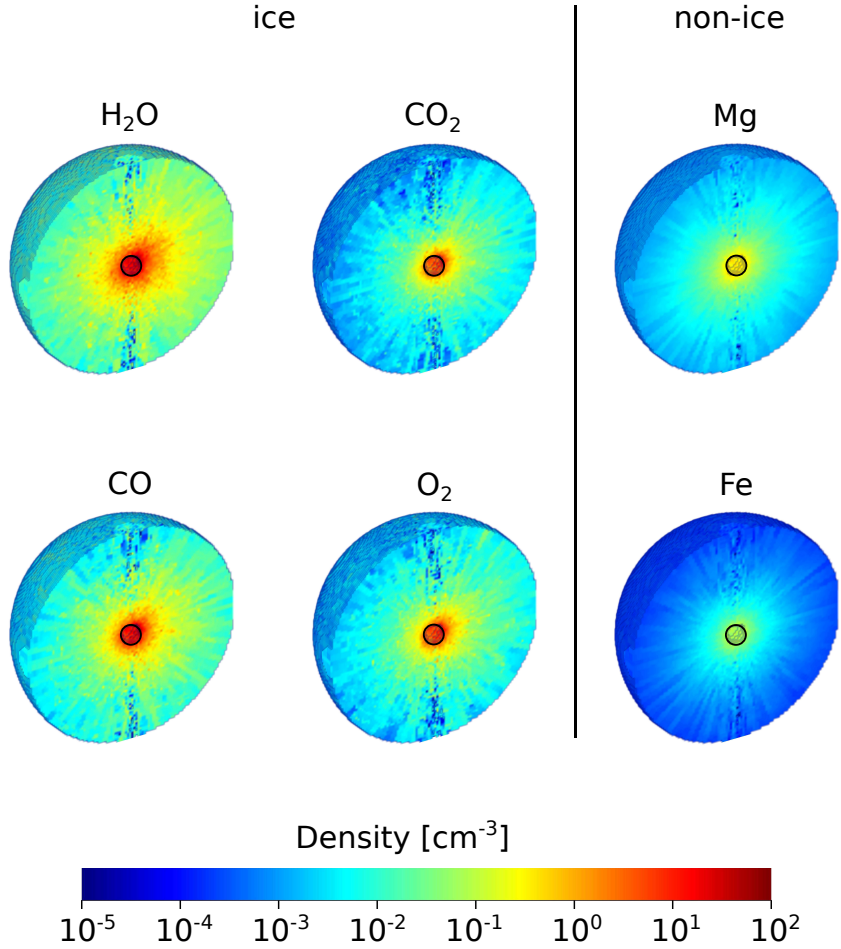}
\caption{Number density cuts in units of cm$^{-3}$ for H$_2$O, CO$_2$, CO, and O$_2$ sputtered from an oxidizing icy surface, Mg sputtered from a LL-type chondritic non-icy surface and Fe sputtered from a CI chondritic non-icy surface for the \textit{`ionosphere'} case. The magnetospheric ram side is on the right hand side, and -\textbf{\textit{B}}$_0$ points upwards. The simulation domain extends between the surface and an altitude of 25000 km or 10.37 Callisto radii.}
\end{center}
\end{figure}

\section{PEP/NIM Measurements}
JUICE, ESA's first Jupiter-dedicated mission, is planned for launch in 2022 and will arrive at Jupiter in 2029, where it will spend at least three years performing detailed observations of Jupiter and its largest icy moons. Several Callisto flybys will provide the JUICE scientific instruments with the opportunity to closely investigate Jupiter's second-largest moon. NIM is a neutral gas and ionospheric ion mass spectrometer designed to investigate the chemical composition of the exospheres of the Galilean moons. The instrument is characterized by a high mass resolution of $M/\Delta M >$ 1100 and an unprecedented sensitivity with a detection level of $1 \times 10^{-16}$ mbar ($\sim$ 1 cm$^{-3}$) for a 5-second accumulation time.

With its high sensitivity, NIM will be able to detect all water-ice related species modeled herein, with H$_2$O surpassing NIM's instrument background as far out as 10$^4$~km above the surface. CO and O$_2$ will become measurable at a few thousand kilometers, whereas the least abundant species modeled, CO$_2$, will be detectable at $\approx$ 10$^3$~km. Determination of the amount of CO, CO$_2$, and CH$_4$ (a molecule highly abundant in the reducing case) in Callisto's exosphere during a Callisto flyby is thus well feasible. With the exosphere representing the surface composition more or less stoichiometrically \citep{Plainaki2018}, knowledge of the chemical composition of Callisto's exosphere will allow us to directly deduce the chemical composition of Callisto's surface and will thus help us to gain insight into Callisto's formation scenario.

Since the non-ice related species are much less abundant than the ice-related species, they pose a challenge concerning measurability. In the study presented herein, out of the two minerals investigated only Mg surpasses NIM's instrument background, and only at closest approach. The detection limit for a given species depends on the orbital characteristics (flyby speed and altitude), possible integration time, and the scale height of the sputtered particles. Since the neutral particle densities and scale heights are dependent on plasma conditions (more energetic plasma ions impose more energy onto the surface particles and higher plasma fluxes result in higher neutral fluxes), more favorable measurement conditions (but also less favorable measurement conditions) can be expected. In fact, the profiles presented herein correspond to an average ion particle density of 0.1~$cm^{-3}$, whereas, according to \citet{Kivelson2004}, the ion particle density ranges from 0.01~$cm^{-3}$ to 0.5~$cm^{-3}$, and the ion bulk velocity ranges from 122~km/s to 272~km/s while 192~km/s was used herein.

\section{Conclusion}
The release of various surface elements caused by plasma sputtering from an assumed icy and a non-icy surface has been studied. The results of the 3D plasma--planetary interaction hybrid model are used as input for a sputter-3D exosphere code that has been used to derive exosphere density profiles and distributions as a function of Callisto's latitude and longitude. We have analyzed two different scenarios, one where no ionospheric shielding takes place (\textit{`no ionosphere'} scenario), and one where an ionosphere only lets 0.07\% of the cold plasma, but all of the hot plasma, reach Callisto's surface (\textit{`ionosphere'} scenario). In the second scenario, due to the associated cold plasma retardation, no cold plasma ion sputtering takes place. These particles are, though, still available for interaction loss processes. The \textit{`no ionosphere'} density profiles show a clear variation with longitude and latitude, with most particles being released in the ram direction (direction of cold plasma arrival). The \textit{`ionosphere'} density profiles are almost uniform, and about 2.5 times lower than the \textit{`no ionosphere'} density profiles. The non-icy density profiles are lower than the icy density profiles by a factor of $\approx$~10 to 100, but exhibit the same geometric characteristics as the icy density profiles. 

NIM will be able detect most major ice and the most abundant non-ice surface constituents. Ice species will be detectable as early as 10$^4$~km above the surface, whereas non-ice species will only be revealed close to the closest approach (200~km). Measurements of the chemical composition of Callisto's exosphere are directly applicable to Callisto's surface, and will shed light onto Callisto's formation scenario as well as provide information on the building blocks of the Jovian system.

\acknowledgments
Data used to create Figures 1--5 presented in this paper can be retrieved from \citet{Vorburger2019}. 

A. Vorburger, A. Galli, M. Rubin, and P. Wurz gratefully acknowledge the financial support by the Swiss National Science Foundation. This work was conducted using resources provided by the Swedish National Infrastructure for Computing (SNIC) at the High Performance Computing Center North (HPC2N), Ume\aa\ University, Sweden. The hybrid model used in this work was in part developed by the DOE NNSA-ASC OASCR Flash Center at the University of Chicago. Jesper Lindkvist is funded by the Swedish National Space Board (SNSB). M. Pfleger, H. Lammer and H. I. M. Lichtenegger acknowledge the support by the Austrian Research Foundation FWF NFN sub-project S11607-N16 ``Particle/Radiative Interactions with Upper Atmospheres of Planetary Bodies Under Extreme Stellar Conditions''.


\end{document}